\documentclass{iau}

\usepackage{amsmath}
\usepackage{graphicx}
\usepackage{xcolor}
\usepackage[hyphens]{url}
\usepackage{hyperref}
\hypersetup{
     colorlinks  = true,
     linkcolor   = blue, 
     anchorcolor = BrickRed,
     citecolor   = blue,
     filecolor   = blue,
     menucolor   = blue,
     runcolor    = cyan,
     urlcolor    = blue
}
\usepackage[authoryear]{natbib}
\usepackage{journals}

\pubyear{2019}
\volume{355} 
\jname{The Realm of the Low-Surface-Brightness Universe}
\editors{D. Valls-Gabaud, I. Trujillo \& S. Okamoto, eds.}

\title[Linking UDGs with dEs] 
{Evolutionary link between ultra-diffuse galaxies and dwarf early-type galaxies.}

\author[Chilingarian et al.]   
{Igor V. Chilingarian$^{1,2}$, Kirill A. Grishin$^{2,3}$, Anton V. Afanasiev$^{2}$, Daniel Fabricant$^{1}$ and Sean Moran$^{1}$}

\affiliation{
$^1$Harvard-Smithsonian Center for Astrophysics,\\
60 Garden St. MS09, Cambridge MA 02138, USA; email: {\tt igor.chilingarian@cfa.harvard.edu}  \\[\affilskip] 
$^2$Sternberg Astronomical Institute, M.V. Lomonosov Moscow State University, \\ 13 Universitetski prospect, Moscow, 119234, Russia\\[\affilskip]
$^3$ Department of Physics, M.V. Lomonosov Moscow State University,\\
Leninskie Gory 1, Moscow, 119234, Russia  \\[\affilskip] 
}

\begin{document}

\maketitle

\begin{abstract}
Spectroscopic studies of low-luminosity early-type galaxies are essential to understand their origin and evolution but remain challenging because of low surface brightness levels. We describe an observational campaign with the new high-throughput Binospec spectrograph at the 6.5-m MMT. It targets a representative sample of dwarf elliptical (dE), ultra-diffuse (UDG), and dwarf spheroidal (dSph) galaxies. We outline our data analysis approach that features (i) a full spectrophotometric fitting to derive internal kinematics and star formation histories of galaxies; (ii) two-dimensional light profile decomposition; (iii) Jeans anisotropic modelling to assess their internal dynamics and dark matter content. We present first results for 9 UDGs in the Coma cluster and a nearby dSph galaxy, which suggest that a combination of internal (supernovae feedback) and environmental (ram-pressure stripping, interactions) processes can explain observed properties of UDGs and, therefore, establish an evolutionary link between UDGs, dSph, and dE galaxies.
\end{abstract}

\firstsection 

\section{Introduction}
Low-luminosity ($M_B>-18$~mag) quiescent galaxies are ubiquitous in galaxy clusters and groups \citep{SB84,FS88} but their origin and evolution still remains a matter of debate. Environmental phenomena such as ram pressure stripping by hot intracluster gas \citep{1972ApJ...176....1G} or frequent tidal interactions with neighbors \citep{1996Natur.379..613M} and internal processes \citep[e.g. supernovae feedback][]{1986ApJ...303...39D} can expel gas from a star-forming dwarf galaxy and turn it into a dwarf elliptical (dE). A sub-class of dwarf quiescent galaxies having enormous spatial extent up-to 10~kpc but low total stellar masses was identified by \citet{SB84} and then re-discovered 2 decades later by \citet{2015ApJ...804L..26V} and re-branded as ``ultra-diffuse galaxies'' (UDGs). It is still unclear whether UDGs and dEs as well as fainter dwarf spheroidal (dSph) galaxies belong to the same class of a galaxy \citep{2018RNAAS...2a..43C} and share similar evolutionary paths or whether they represent different types with different formation scenarios.

Faint surface brightness levels severely limit our ability to study dwarf early-type galaxies. UDGs are the most extreme case: unlike dSphs they do not exist in the Local Group so we do not have the luxury of resolved stellar population studies. At the same time, we will be able to conclude the dE/UDG/dSph formation debate if we are able to: (i) probe the dark matter content along the luminosity and size range covering these galaxy types; (ii) probe stellar dynamics and measure the degree of rotational support; (iii) assess their stellar population properties and/or star formation histories.

To complement known dE/UDG samples, one can potentially identify ``future dEs/UDGs'' while their stars are still young and the surface brightness is high despite low stellar surface density and total stellar mass. We found 13 such spatially extended, young (stellar age $t<1.5$~Gyr) low-mass galaxies with no current star formation using the Reference Catalog of Spectral Energy Distributions of galaxies \citep[\url{http://rcsed.sai.msu.ru/};][]{RCSED}, 12 of which turned to be members of the massive galaxy clusters Coma and Abell~2147.

Here we describe our observational campaign aimed at studying the links between different classes of low-mass galaxies, present the data analysis approach and first results.

\section{Observational campaign and complementary archival data}

\begin{figure}
\begin{center}
 \includegraphics[width=0.8\textwidth]{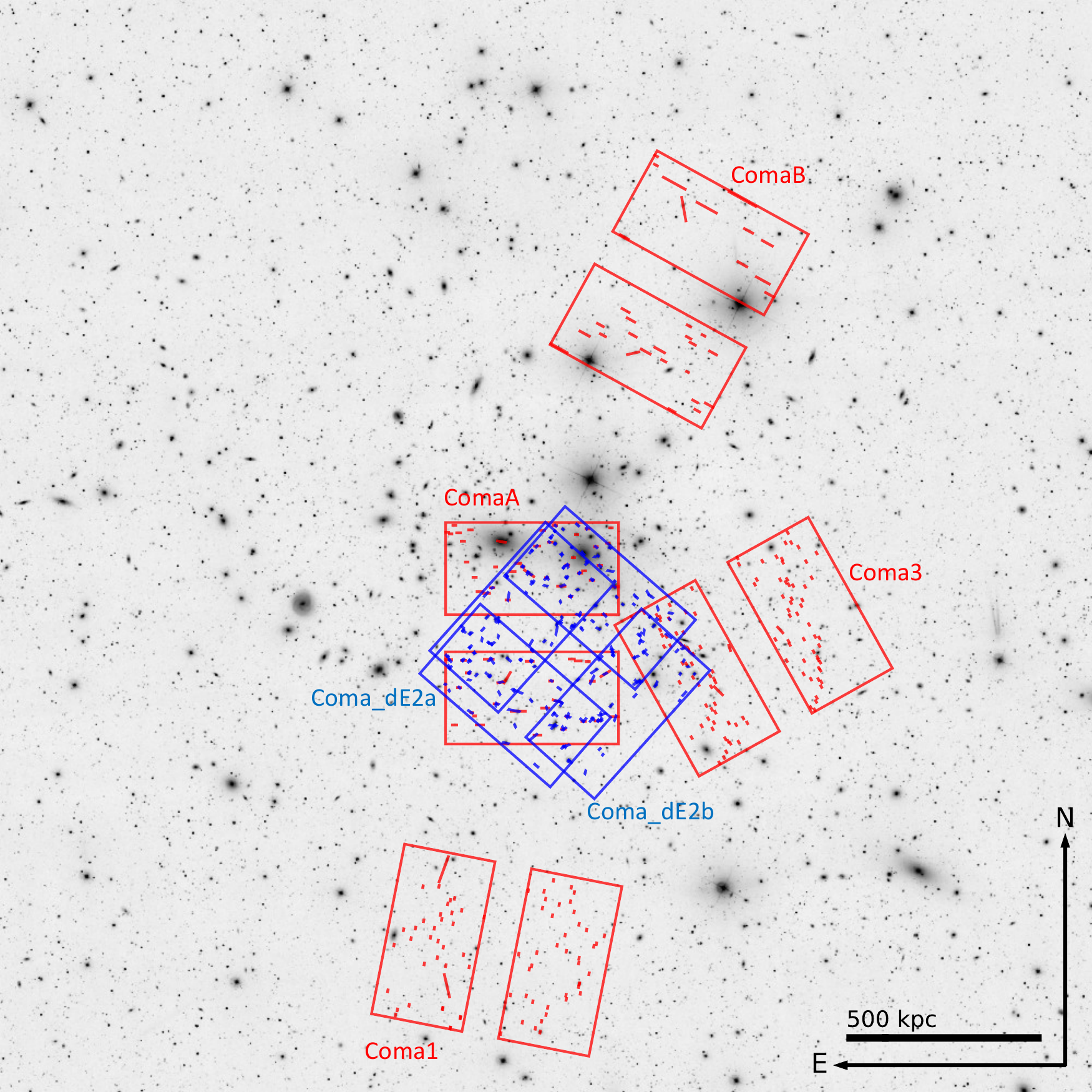} 
 \caption{The layout of Binospec slitmasks observed as of May/2019 in the Coma cluster region. The 4 masks shown in red were observed during commissioning and first months of operations in Dec/2017 -- Jun/2018. The 2 masks shown in blue are located in the area covered by archival HST observations; they were observed in May/2019.}
  \label{fig1_chil}
\end{center}
\end{figure}

To study the UDG--dE--dSph evolutionary connection one needs detailed information on internal dynamics (including dark matter) and stellar content of representative samples of objects of every galaxy type. While for dSphs there is a wealth of literature data presenting resolved stellar population studies of Local Group galaxies \citep[see e.g.][]{McConnachie12}, the situation with UDGs and dEs is far from ideal. 

For our deep spectroscopic follow-up we use Binospec \citep{2019PASP..131g5004F}, a new dual-channel high-throughput multi-object spectrograph operated at the Cassegrain f/5 focus of the 6.5-m converted MMT at Mt.~Hopkins, Arizona. Binospec masks cover two 15$\times$7~arcmin fields of view (450$\times$210~kpc in the Coma cluster) separated by a 1.5~arcmin wide gap and can contain up-to 300 targets on every side of the mask. We modified the slitmask design procedure to allow for tilted slits so that we could obtain major and/or minor axis kinematics for many galaxies at once.

Our original observing program was a part of the commissioning and science verification of Binospec. It targeted 11 diffuse post-starburst ``future UDGs'' in the Coma and Abell~2147 clusters (5 slitsmasks) as primary targets to obtain their spatially resolved major axis kinematics. UDGs and dEs were filler targets and we did not align slits along their major axes. Then we started a dedicated follow-up program in May/2019 by specifically targeting dEs and UDGs in the central part of the Coma cluster \citep{Koda15} where archival Hubble Telescope Images are available. We placed slits along major and minor axes for several dozens of objects. The exposure times were between 2 and 4~h per pointing; we used the 1000~gpm grating providing a spectral resolving power $R=4800$ in the wavelength range between 3800 and 5400~\AA\ with a 0.37~\AA/pix sampling. 

In total, we observed 8 slitmasks: six in the Coma cluster, one in Abell~2147, and one in the M~81 group, where we targeted the dwarf spheroidal galaxy KDG~64 (UGC~5442). It is spatially extended and quite luminous, which makes it the best UDG analog we can find in the very nearby Universe. In Figure~\ref{fig1_chil} we show the layout of Binospec slitmasks in the Coma cluster. In total, we targeted about 150 low-luminosity ($-18.0<M_B<-13.5$~mag) early-type galaxies in Coma and Abell~2147 including about 30 UDGs.

To complement our spectroscopic observations we use imaging in up-to 13 bands that cover ultraviolet (\emph{GALEX FUV}/\emph{NUV}), optical (SDSS/CFHT MegaPrime \emph{ugriz}), and near-infrared (CFHT WIRCAM \emph{J/Ks}; MMT MMIRS \emph{J/Ks}; Magellan FourStar \emph{JHKs}; Spitzer \emph{IRAC 3.6/4.5~$\mu$m}) spectral domains.  \emph{GALEX}, \emph{Spitzer}, \emph{CFHT}, and \emph{SDSS} datasets were retrieved from archives, while 
FourStar \citep{Persson+13} and MMIRS \citep{McLeod+12} imaging was a part of our own observational program. With FourStar and MMIRS we used on-source exposure times of 8--12~min per filter split into short 10--30~sec dithered exposures. We reduced FourStar and MMIRS imaging using the {\sc fsred} and MMIRS data reduction pipelines \citep{Chilingarian+15} respectively.

\section{Data analysis}

\begin{figure}
\begin{center}
 \includegraphics[width=1.0\textwidth]{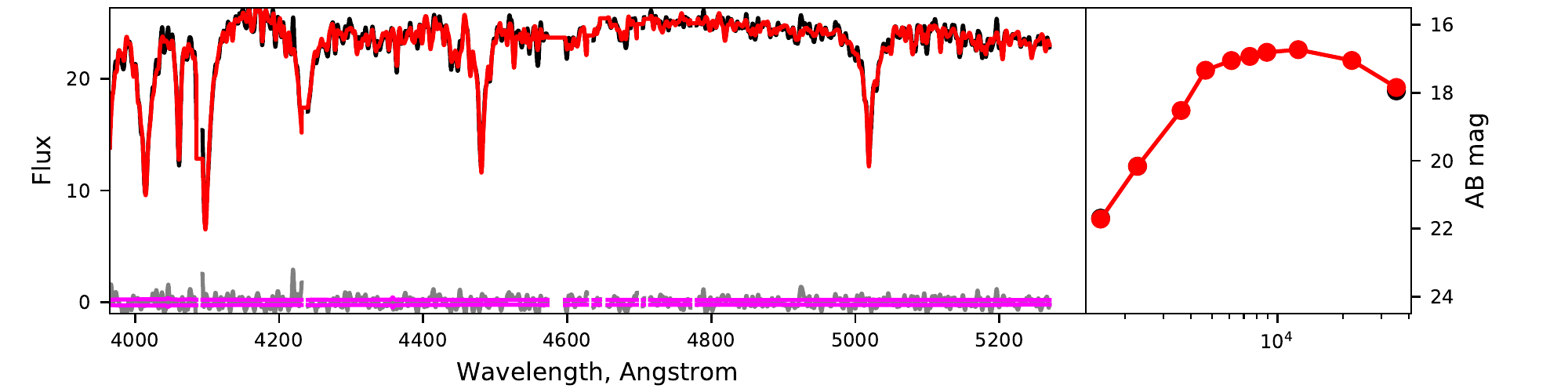}\\
 \includegraphics[width=0.8\hsize]{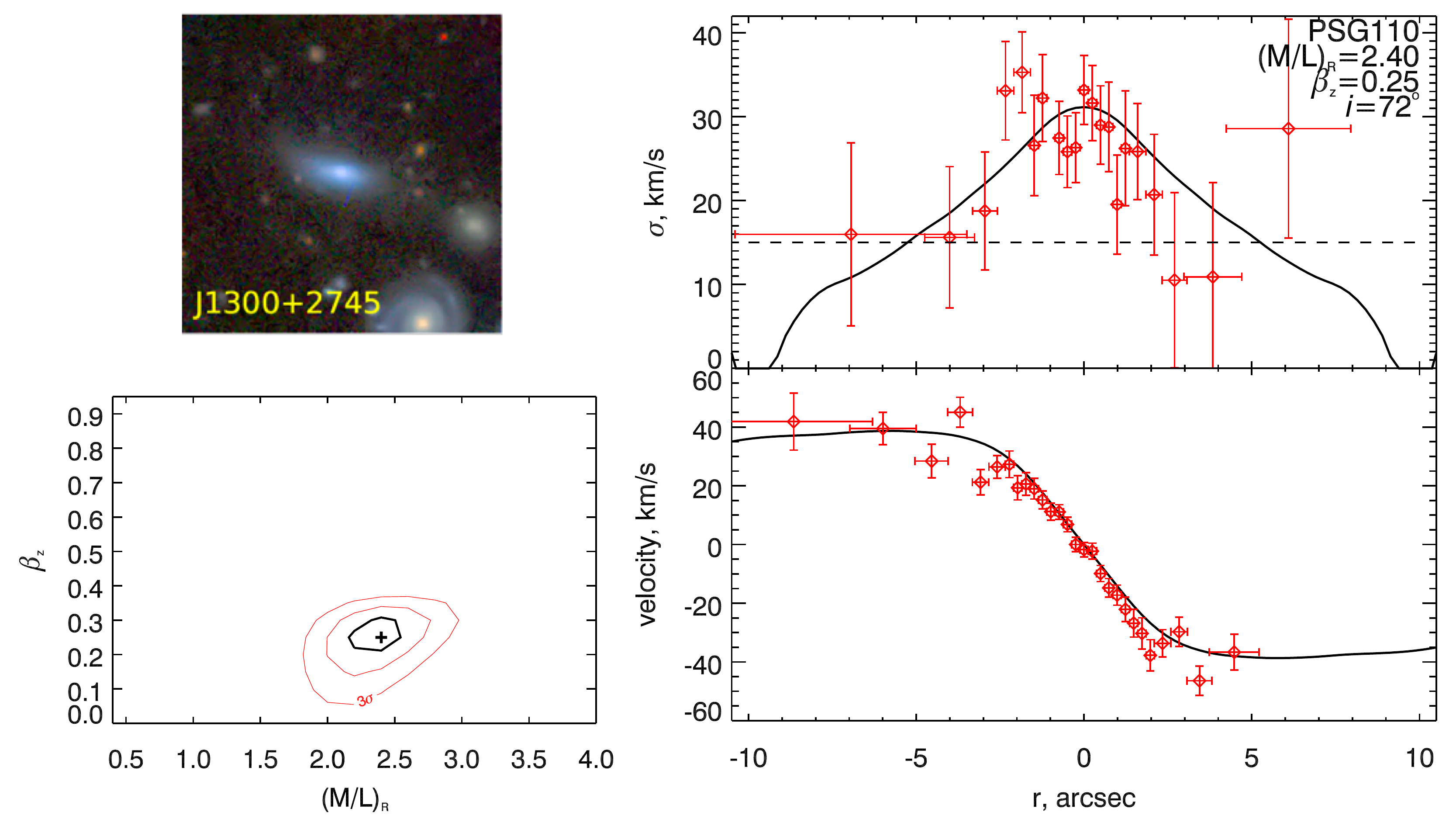}
 \caption{Results of our data analysis applied to one of the ``future UDGs'' in the Coma cluster. Top row: the results of the {\sc NBursts+phot} spectrophotometric analysis (data and models are shown in black and red; Binospec spectrum on the left and SED on the right); middle row: an optical color image of a galaxy on the left; the stellar velocity dispersion profile (red) and its best-fitting Jeans model (black) on the right; bottom row: a map of confidence levels for the determination of the dynamical mass-to-light ratio and orbital anisotropy on the left and the stellar radial velocity profile (red) and its best-fitting Jeans model (black) on the right.}
  \label{fig2_chil}
\end{center}
\end{figure}

For every galaxy we need to obtain the following two key characteristics: (i) star formation history (SFH) and stellar mass-to-light ratio; (ii) internal kinematics and dynamical mass-to-light ratio. Then we will be able to measure the dark matter content and the degree of rotational support that will allow us to see which formation scenario agrees best with observations for that particular object.

Our data analysis includes three main components. (i) We fit science-ready Binospec spectra produced by the data reduction pipeline \citep{2019PASP..131g5005K} simultaneously with multi-wavelength broad-band $k$-corrected \citep{CMZ10,CZ12} photometric measurements against several grids of stellar population models using the {\sc NBursts+phot} spectrophotometric fitting technique \citep{Nburstsphot}, which recovers a parametric SFH and internal kinematics. We use {\sc miles} \citep{2015MNRAS.449.1177V} and {\sc pegase.hr} \citep{LeBorgne+04} simple stellar populations and more advanced models described in a companion paper \citep{PSGM_IAU355}. We use adaptive binning along the slit to reach sufficient signal-to-noise ratios for the analysis of kinematics (3--5 per bin per pixel) and stellar populations (7--12 per bin per pixel) (ii) We fit optical and near-infrared images using the {\sc galfit} code \citep{Peng10} with one or two S\'ersic components and then run their parameters through an {\sc idl} script described in \citet{Afanasiev+18} to perform a multiple Gaussian expansion (MGE). (iii) Then we feed the internal kinematics and MGE of a brightness profile to the Jeans Anisotropic Modelling (JAM) code \citep{Cappellari08} and derive the dynamical mass-to-light ratio and the radial anisotropy of stellar orbits in a galaxy.

In Figure~\ref{fig2_chil} we present an example of our data analysis approach applied to one of the 10 ``future UDGs'' in the Coma cluster. Thanks to deep observations and high surface brightness of the galaxy, we can reach as far as 3--4~kpc (2--2.5 half-light radii) in radial velocity and velocity dispersion profiles. That allows us to break the degeneracy between the mass-to-light ratio $(M/L)_{R,\mathrm{dyn}}=2.4\pm0.1$ and orbital anisotropy $\beta=0.22\pm0.08$. Combined with the stellar mass-to-light ratio $(M/L)_{R*}=0.39\pm0.06$ it yields the dark matter fraction within 2~$r_e$ of about 83$\pm$6~\%.

\section{First results}

\begin{figure}
    \centering
    \includegraphics[width=\hsize]{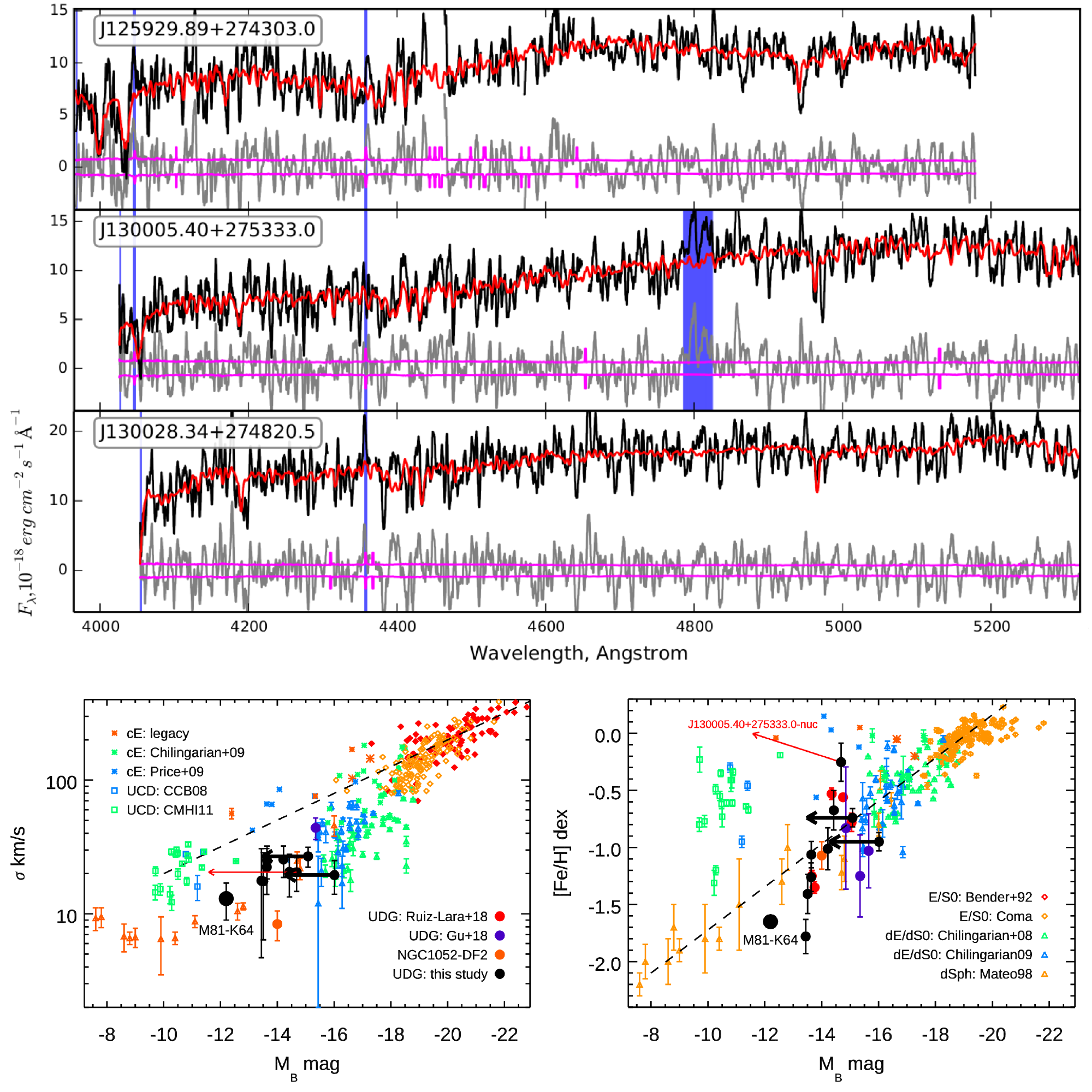}
    \caption{Top panels: examples of 3 spectra of Coma cluster UDGs collected with Binospec, their best-fitting templates and fitting residuals. Bottom panels: Faber--Jackson (left) and mass--metallicity (right) relations for early-type galaxies and compact stellar systems, the updated version of figure~3 from \citep{Chilingarian+19} where we included measurements for KDG~64.}
    \label{fig3_chil}
\end{figure}

The first results that we obtained from our observational campaign are stellar population and internal dynamics of 9 UDGs in the Coma cluster which prove their link to dwarf early-type galaxies \citep{Chilingarian+19}. This sample includes 7 UDGs from the \citet{Koda15} UDG catalog and 2 younger ``future UDG'' galaxies identified photometrically, which resemble UDGs in terms of stellar surface density but are not as young and bright as the 13 galaxies selected from RCSED mentioned earlier. Despite relatively low signal-to-noise ratios (see spectra in the top part of Figure~\ref{fig3_chil}), the data quality allowed us to detect radial velocity gradients using 4--7 measurements along the slit and obtain integrated velocity dispersion measurements and mean stellar population characteristics for 6 galaxies, while for the remaining 3 we obtained spatially resolved velocity dispersion profiles. 

Even in this limited sample of only 9 galaxies we see a great diversity of observed properties: three of them exhibit major axis rotation, two others have highly anisotropic stellar orbits, and one either has a triaxial intrinsic shape or is currently undergoing a tidal interaction with a neighbor (it also hosts a luminous metal-rich intermediate-age nuclear star cluster). This situation resembles the sub-classes of dwarf elliptical galaxies in the Virgo cluster identified by \citet{2007ApJ...660.1186L} and, in fact, every one of the 9 UDGs can be assigned to one of the dE sub-classes. The estimated dark matter fractions lay in the range between 50 and 90~\%\ slightly exceeding the typical values in brighter cluster dEs.

We placed these 9 UDGs on the \citet{1976ApJ...204..668F} and mass--metallicity relations shown in the bottom part of Figure~\ref{fig3_chil} and compared them to the available literature data for giant and dwarf early-type and dSph galaxies \citep{1992ApJ...399..462B,Chilingarian+08,Chilingarian09,Mateo98}, compact stellar systems \citep{Chilingarian+09,Price+09,CCB08,CMHI11} and a few UDGs \citep{2018ApJ...859...37G,2018MNRAS.478.2034R,Danieli+19}. The 7 UDGs and 2 ``future UDGs'', if we predict their positions on the plots after 5~Gyr of passive evolution (see black arrows in Figure~\ref{fig3_chil}), fill the gap between dSph and dE galaxies, thus providing an additional argument for the evolutionary link between these classes. On the other hand, compact stellar systems (ultra-compact dwarf galaxies and more luminous compact ellipticals), which are known to form differently via tidal stripping of massive progenitors form a parallel sequence on the Faber--Jackson relation at 2.5 higher values of velocity dispersions. The compact nucleus of one of the UDGs (end of the red arrow in Figure~\ref{fig3_chil}) is located in the middle of the UCD loci on both relations, which might suggest that this galaxy is not a UDG but rather an ``under-stripped'' UCD.

In Figure~\ref{fig3_chil} we also display KDG~64 (Afanasiev et al. in prep.), a luminous dSph in the M~81 group, which exceeds by its size and luminosity all dwarf spheroidal satellites of the Milky Way and Andromeda galaxy and is almost as large as the Local Group dwarf ellipticals NGC~185 and NGC~147. It follows the same trends on both relations established by dwarf and giant early-type galaxies. 

Our early results, therefore, suggest that the standard evolutionary paths of dEs (ram pressure stripping, galaxy harassment, supernovae feedback) can and probably should be applied to UDGs to explain their observed properties.

\acknowledgements
IC, KG, and AA acknowledge the Russian Science Foundation grant 19-12-00281.

\bibliographystyle{aasjournal}
\bibliography{UDG}

\end{document}